\documentclass[twocolumn,superscriptaddress,aps,10pt]{revtex4}
\usepackage{amsfonts}
\usepackage{amsmath}
\usepackage{amssymb}
\usepackage{graphicx}
\usepackage{dcolumn}
\usepackage{times}
\usepackage{color}

\DeclareMathOperator{\sech}{sech}

\begin{document}

\title{\bf
Dark solitons and vortices in $\mathcal{PT}$-symmetric nonlinear media:\\
from spontaneous symmetry breaking to nonlinear $\mathcal{PT}$ phase transitions}
\author{V.~Achilleos}
\affiliation{Department of Physics, University of Athens, Panepistimiopolis,
Zografos, Athens 157 84, Greece}
\author{P.G.~Kevrekidis}
\affiliation{Department of Mathematics and Statistics, University of Massachusetts,
Amherst, Massachusetts 01003-4515, USA}
\author{D.J.~Frantzeskakis}
\affiliation{Department of Physics, University of Athens, Panepistimiopolis,
Zografos, Athens 157 84, Greece}
\author{R.\ Carretero-Gonz\'alez}
\affiliation{Nonlinear Dynamical Systems Group,
Computational Sciences Research Center, and Department of Mathematics and
Statistics, San Diego State University, San Diego, CA 92182-7720, USA}


\begin{abstract}

We consider nonlinear analogues of Parity-Time
($\mathcal{PT}$) symmetric linear systems
exhibiting defocusing nonlinearities.
We study the ground state and excited
states (dark solitons and vortices)
of the system and report the following remarkable features.
For relatively weak values of the parameter $\varepsilon$ controlling the strength
of the $\mathcal{PT}$-symmetric potential, excited states undergo
(analytically tractable) spontaneous symmetry breaking;
as $\varepsilon$
is further increased, the ground state and first excited state,
as well as branches of higher multi-soliton (multi-vortex) states,
collide in pairs and disappear in blue-sky bifurcations, in a way
which is strongly reminiscent of the linear $\mathcal{PT}$-phase transition
---thus termed the nonlinear $\mathcal{PT}$-phase transition.
Past this critical point, initialization of, e.g., the former ground state
leads to spontaneously emerging solitons and vortices.

\end{abstract}

\maketitle

\section{Introduction}

Over the past decade, and since its original inception
\cite{bend1,bend2}, the theme of $\mathcal{PT}$-symmetric Hamiltonians
has gained considerable momentum in the physics and applied mathematics
communities.
Such systems,
respecting both
Parity ($\mathcal{P}$)
and Time-reversal ($\mathcal{T}$) symmetries
---still exhibiting
real spectra while non-Hermitian--- provided an intriguing alternative
to standard Hermitian quantum mechanics.
Note that for
a standard Schr{\"o}dinger type Hamiltonian with a generally complex
potential $U$, the $\mathcal{PT}$ symmetry dictates that the potential satisfies
the condition $U(x)=U^{\ast}(-x)$ [where $(\cdot)^{\ast}$ stands for
complex conjugation].

Despite the theoretical
appeal of such models,
it was only recently shown \cite{christo1}
that optics could be an ideal playground for the physical/experimental realization
of systems featuring the $\mathcal{PT}$ symmetry.
However, this also added another element in the
interplay, namely nonlinearity. In that context, the considerations
of Ref.~\cite{christo1} extended from bright and gap solitons
to linear (Floquet-Bloch) eigenmodes in periodic potentials, examining how these
coherent structures are affected by the genuinely complex, yet
$\mathcal{PT}$-symmetric potentials. More recently, experimental results were reported
both in nonlinear optical systems \cite{salamo,kip}
and electronic analogs thereof~\cite{tsampikos_recent}. These, in turn, have triggered
a wide range of theoretical studies on
nonlinear lattices with either linear ~\cite{kot1,sukh1,kot2,grae1,grae2,kot3,pgk,dmitriev1,dmitriev2}
or nonlinear~\cite{miron,konorecent,konorecent2} $\mathcal{PT}$-symmetric
potentials and, more recently, on harmonic $\mathcal{PT}$-symmetric potentials \cite{vvk}.

While the above volume of work has examined numerous features
extending from bright solitons to defect modes, and from gap solitons
to $\mathcal{PT}$-lattices, the consideration of defocusing nonlinearities,
and especially of dark solitons
has been extremely limited (see, e.g., Ref.~\cite{Li}).
Little attention (and again
chiefly in the focusing nonlinearity
case~\cite{christo1}) has also been paid to
$\mathcal{PT}$-symmetric systems
in higher-dimensional settings and the corresponding interplay
with nonlinear states such as vortices.

In the present work, we study systems with
$\mathcal{PT}$-symmetric
Hamiltonians exhibiting defocusing
nonlinearities, and focus on
the existence, stability and dynamical properties of the ground state and excited states, i.e.,
dark solitons and
vortices.
Our main findings for a prototypical $\mathcal{PT}$-symmetric potential, which is harmonic in
its real part and has a
localized imaginary part
(parametrized by an amplitude parameter $\varepsilon$)
are summarized as follows:
1) dark solitons
are shown to be subject
to spontaneous symmetry-breaking (SSB) instabilities for small $\varepsilon$;
2) for higher values of $\varepsilon$, the ground state and the
first excited state (single dark soliton),
as well as pairwise ---e.g., 2nd and 3rd, 4th and 5th etc.--- higher excited states
(respective multiple dark soliton states)
are subject to a nonlinear analogue
of the $\mathcal{PT}$-phase transition, colliding
and disappearing in a set of blue-sky bifurcations;
3) beyond this critical point, the system acts as a soliton generator,
spontaneously emitting dark multi-soliton structures.
4) {\it All} of these
features have direct counterparts for vortices in two-dimensional
settings, illustrating the generic nature of these findings.
%

The paper is organized as follows. In Sec.~II we introduce the model and study its one-dimensional 
(1D) version; in the same section, analytical and numerical results for the statics and dynamics 
of the ground state and dark solitons are presented; we also briefly touch
upon the potential effects of noise in the gain profile. In Sec.~III, we discuss nonlinear 
$\mathcal{PT}$-phase transitions occurring in the 1D setting and study the dynamics of the system 
beyond the relevant critical point. In Sec.~IV, we generalize our findings in the two-dimensional (2D) 
setting, studying vortex states and their nonlinear $\mathcal{PT}$-phase transitions. Finally, in Sec.~V,  
we present a summary of our results. 

\section{Model,
Ground State and Dark Solitons}
Our model,
motivated by the above nonlinear optical considerations
(but also by ones pertinent to nonlinear phenomena in
Bose-Einstein condensates (BECs) \cite{emergent}), will be,
for the 1D 
setting, as follows:
\begin{eqnarray}
i \partial_t u=-\frac{1}{2} \partial_x^2 u + |u|^2 u + [V(x) + i W(x)] u,
\label{PT1}
\end{eqnarray}
where $u$ is the complex electric field envelope
(or the macroscopic wavefunction in BECs), $t$ denotes the
propagation distance (or time in BECs) and $x$ is the transverse direction.
For a $\mathcal{PT}$-symmetric Hamiltonian, the real and imaginary parts
of the potential must satisfy $V(x)=V(-x)$ and $W(x)=-W(-x)$. Below we focus on the
case of a real parabolic potential
$$V(x)=\frac{1}{2}\Omega^2 x^2,$$
modeling the transverse distribution of the refractive index
(or the external trap in BECs), while the imaginary part $W(x)$ is considered to be an odd,
localized function
of spatial width $\ll \Omega^{-1}$; as a prototypical
example, we will consider 
\begin{equation}
W(x)=\varepsilon x e^{-x^2/2}.
\label{W}
\end{equation}
A generalization of this model in two-dimensions will be studied in 
Sec.~\ref{sec:2D}.

Here, it is important to notice that the evolution of the power
(number of atoms in BECs), $N=\int|u|^2 dx$, is governed by the equation
\begin{equation}
dN/dt=2\int |u|^2 W(x) dx. 
\label{dNdt}
\end{equation}
Thus, since $W(x)$ is odd, if $|u|^2$ is even
then $N$ is conserved ($dN/dt=0$). Below, we show that this is the case
for the stationary states of the system that we will consider here, i.e.,
the ground state and excited states (dark solitons in 1D and vortices in 2D).

\subsection{Ground state}

\begin{figure}[tbp]
\includegraphics[width=6.5cm]{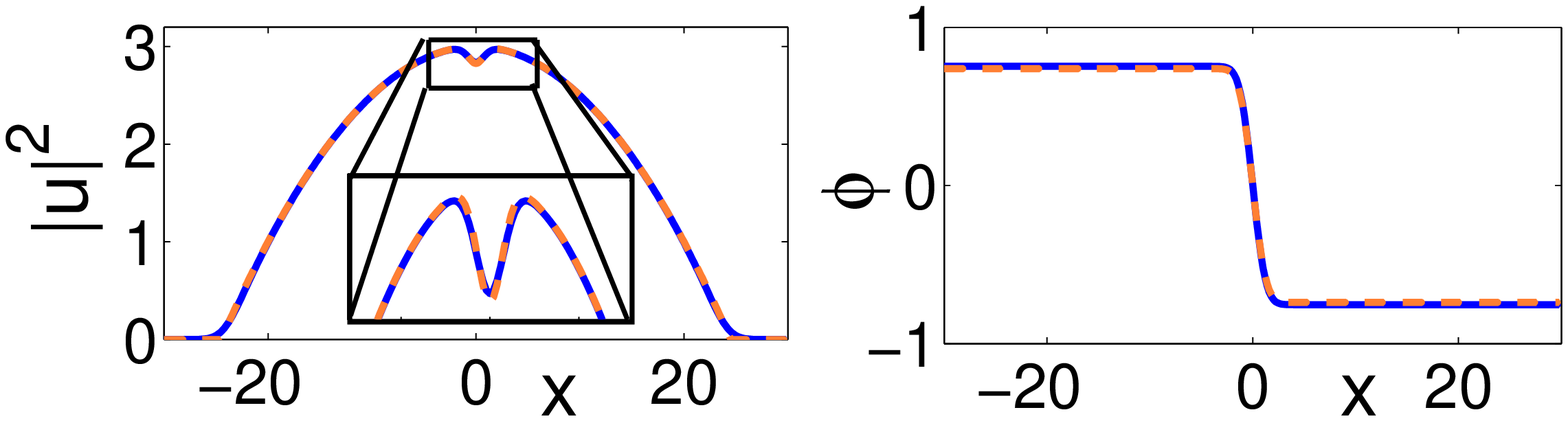}
\includegraphics[width=5.5cm]{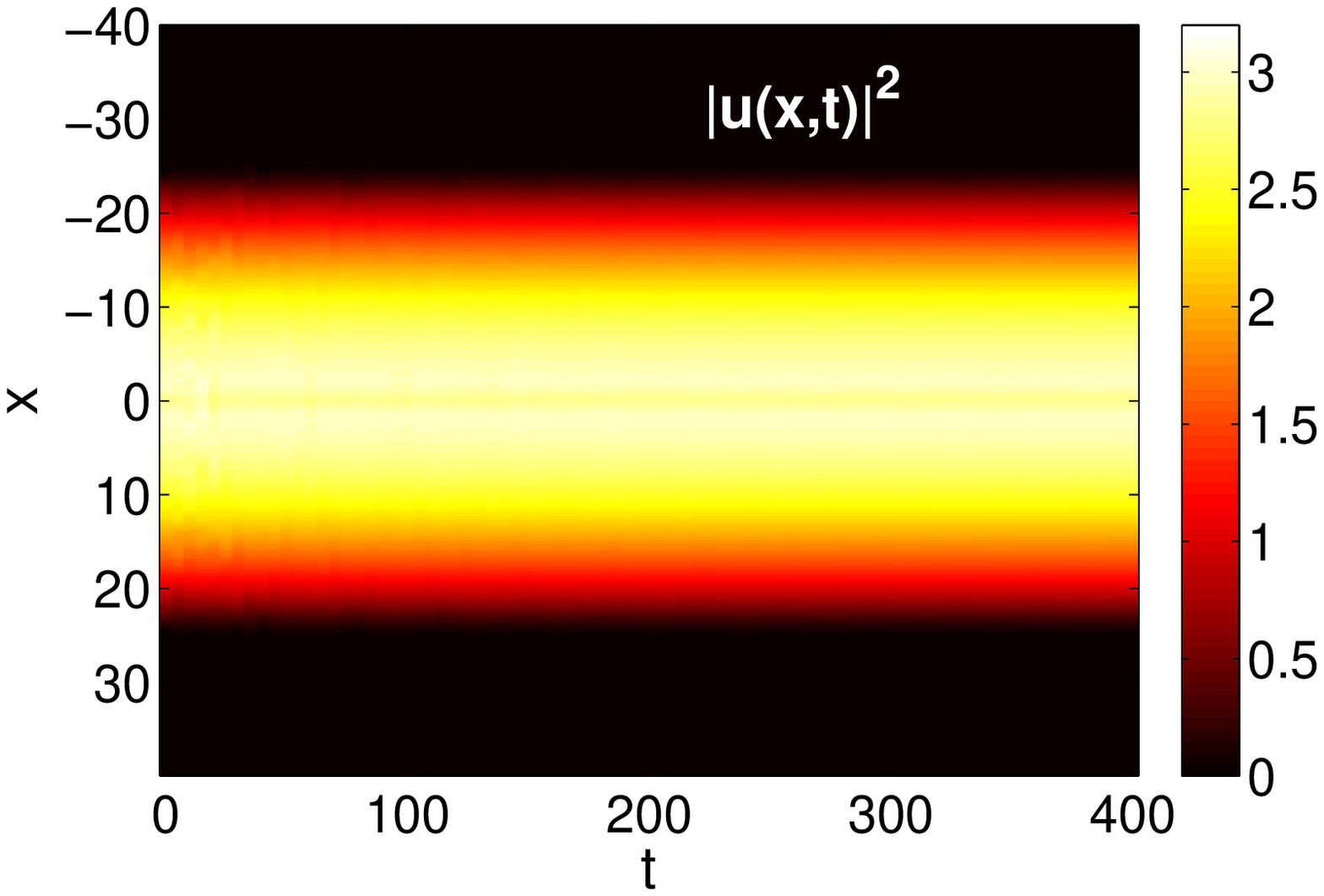}
\caption{(Color online)
Top panels: density (left) and phase (right) of the numerically obtained TF background
[solid (blue) line] compared to the prediction of Eqs.~(\ref{tf23})
[dashed (orange) line]; the inset shows the characteristic density dip
induced by $W(x)$ at the origin.
%
Bottom
panel: contour plot showing the evolution of the density $|u(x,t)|^2$ with an initial condition $u(x,0)=u_b(x)$.
Parameter values are: $\mu=3$, $\Omega=0.1$ and $\varepsilon=0.4$.}
\label{fig0}
\end{figure}

We first analyze the most fundamental state, namely the ground state
of the system, shown in Fig.~\ref{fig0}.
We
seek stationary solutions of Eq.~(\ref{PT1}) in the form
$u= u_b(x) \exp(-i\mu t)$, where $\mu$ is the propagation constant (or the chemical potential in
BECs), while
the background field $u_b$ obeys the equation:
\begin{eqnarray}
-\frac{1}{2}\partial_x^2 u_{b} + |u_b|^2 u_b + [V(x) + i W(x)]u_b-\mu u_b=0.
\label{PT2}
\end{eqnarray}
For a sufficiently small imaginary potential, $W(x)=\varepsilon\tilde{W}(x)$ [with ${\rm max} \{|\tilde{W}(x)|\} = O(1)$],
where $\varepsilon \ll 1$ \cite{note1},
and when the inverse width $\Omega^{-1}$ of $V(x)$ is sufficiently large, $\Omega \sim \varepsilon $,
we may find  ---in the Thomas-Fermi (TF) limit--- an approximate solution of
Eq.~(\ref{PT1}).
This is of the form
%
\begin{equation}
u_b=\left[\sqrt{\mu}+f\left(x\right)\right]\exp[i\phi(x)],
\label{TF}
\end{equation}
%
%
where the amplitude and phase
$f(x)$ and $\phi(x)$
(considered to be small, of order $\mathcal{O}(\varepsilon^2)$ and $\mathcal{O}(\varepsilon)$, respectively) are given by:
%
%
%
%
\begin{equation}
f(x)=-\frac{1}{2\sqrt{\mu}} \left(V +2 {\cal W}^2\right),
\quad
\phi(x)= 2\int {\cal W}\, dx.
\label{tf23}
\end{equation}
%
where ${\cal W}=\int W dx$
(note that above integrals are indefinite ones).
Contrary to the conservative case ($\varepsilon=0$) \cite{emergent}, this
TF background
is characterized by a density dip located at the center ($x=0$) and
a nontrivial phase distribution. Both features are shown in the top panels of Fig.~\ref{fig0}, where
the analytical result is compared with the numerical one, which was obtained by means of
a fixed-point algorithm (Newton's method). It is observed that the agreement between
the two is excellent.
%
%
%
%
Furthermore, a linear stability ---Bogoliubov-de Gennes (BdG)--- analysis (see, e.g., Ref.~\cite{review})
shows that the background $u_b(x)$ is stable against small perturbations.

%

The stability of the analytically found ground state solution has also been tested by means of
direct numerical integration of Eq.~(\ref{PT1})
using as an initial condition Eq.~(\ref{TF}), and it has been confirmed that it
remains stable for long times, as shown in the bottom panel of Fig.~\ref{fig0}.
Note that this occurs even for relatively large $\varepsilon$ (e.g., for $\varepsilon=0.4$
used in the figure).
As is observed, the ground state is practically
stationary: it evolves in time only for a small initial time interval, during which the tails of the
analytical solution slowly approach the ones of the
exact ground state solution. This is expected because the spatial structure
of the tails is different in the numerical and approximate solutions ---due to the
cutoff structure of the latter that becomes regularized--- see, e.g., Ch.~6 of Ref.~\cite{Peth}.

Beyond this initial transient (which is chiefly seeded by the tail
and not the core of the wave form), the profile remains practically identical to the exact numerical solution in a wide spatial region ---more than $90\%$ of the
TF radius $\sqrt{2\mu}/\Omega$; in fact, as we will show below,
the relative error
is of order of $\mathcal{O}(10^{-2})$ for $\varepsilon=\mathcal{O}(10^{-1})$.
Thus, indeed, the error is of $\mathcal{O}(\varepsilon^2)$ and stays bounded
within that order of approximation for extremely long evolution times; this indicates that, since the system
under consideration
possesses gain-loss,
this state is indeed an attractor.
This is, naturally, also the case
at the vicinity of the central region where the imaginary potential $W(x)$ acts: our approximate solution
for the ground state, Eqs.~(\ref{TF})-(\ref{tf23}), predicts a localized dip in the density, with an approximation of the order $\mathcal{O}(\varepsilon^2)$.

\begin{figure}[tbp]
\includegraphics[width=6.0cm]{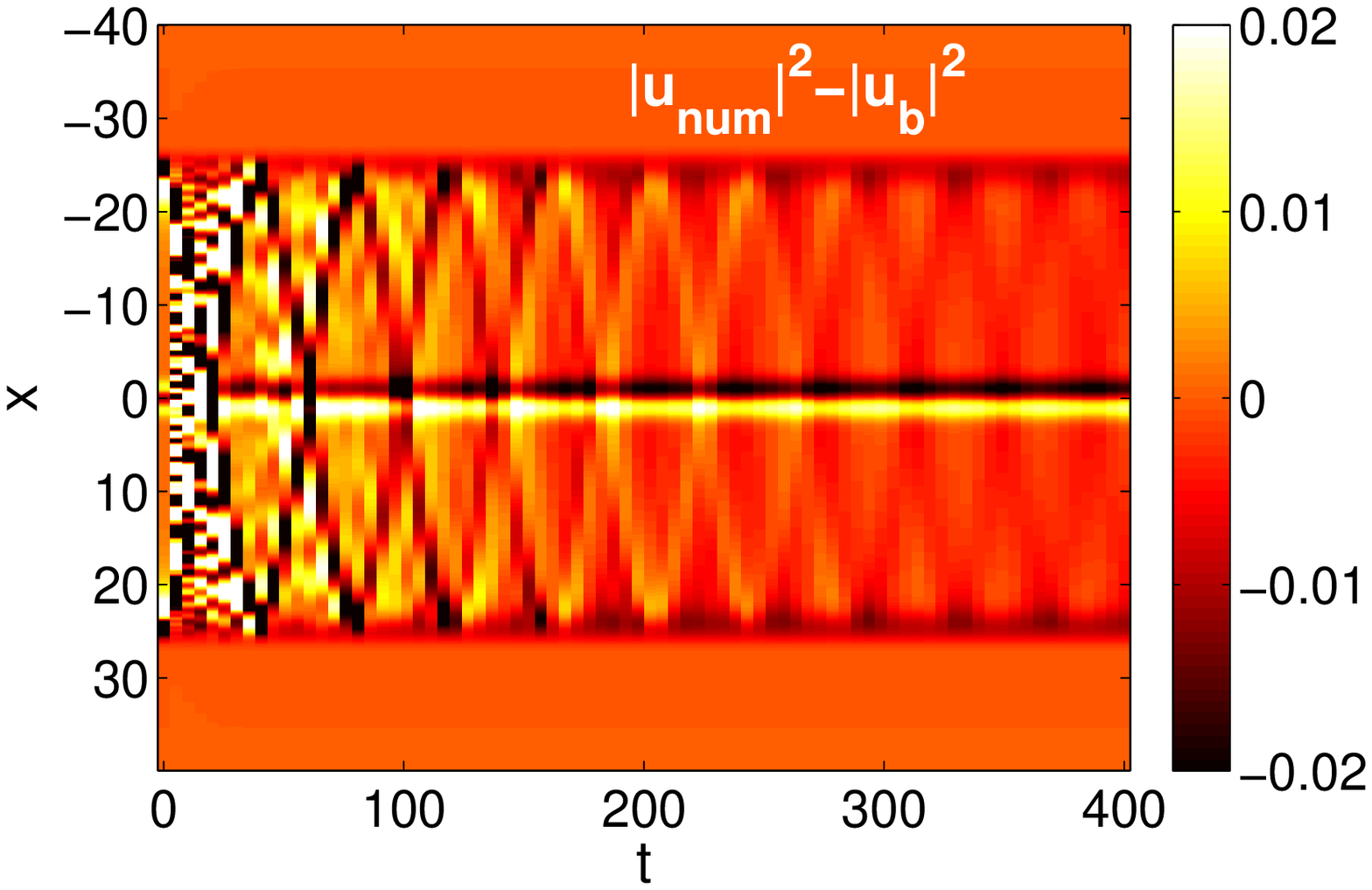}
\includegraphics[width=6.0cm]{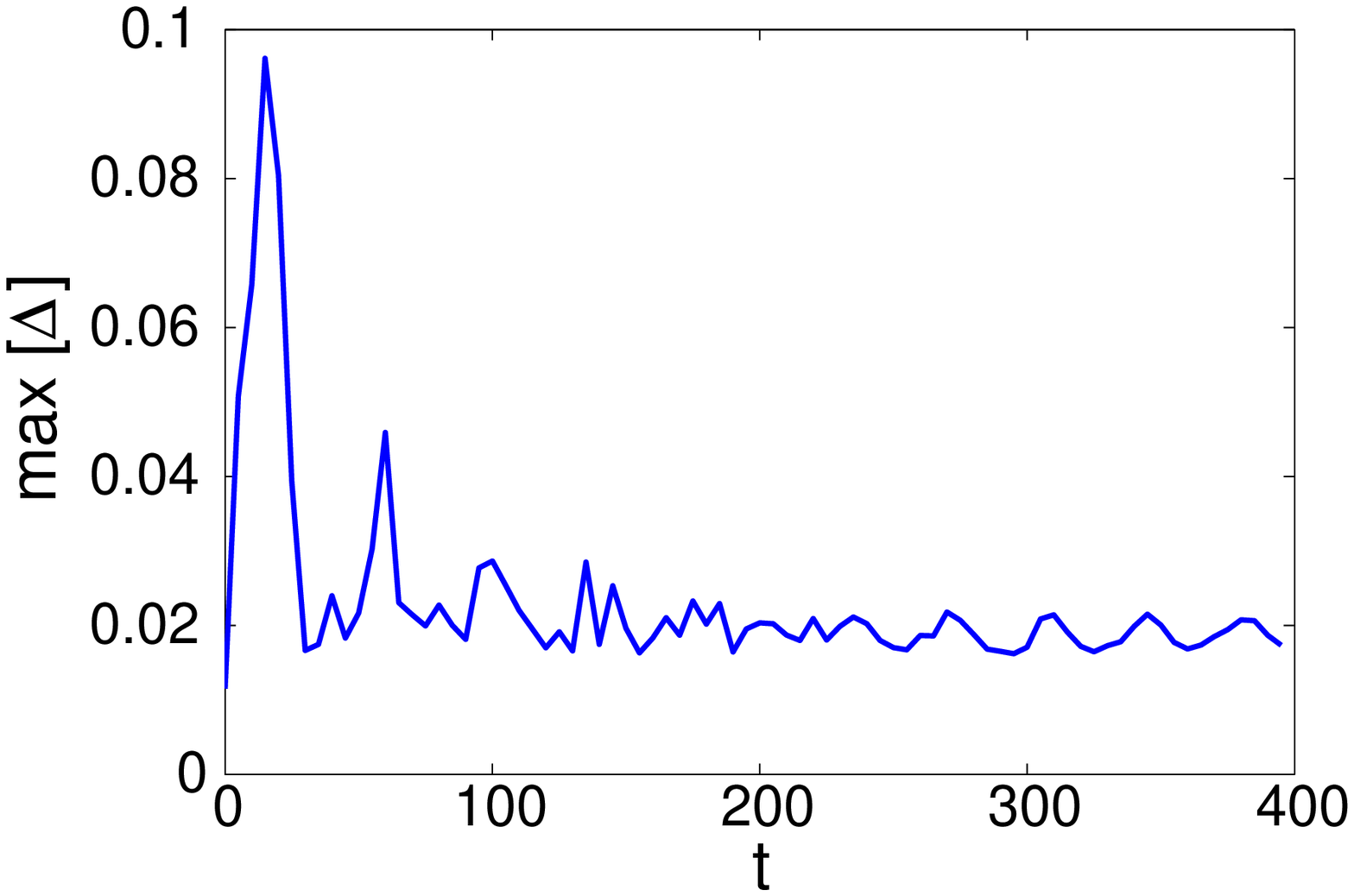}
\caption{(Color online)
Top panel:
%
contour plot showing the evolution of the density difference $\Delta$ [cf. Eq.~\ref{D}]
as a function of time.
Bottom panel: the maximum value of $\Delta$ at the center.
Parameter values are the same as the ones in Fig.~\ref{fig0}.
}
\label{1}
\end{figure}
%

The validity of
the above arguments is clearly illustrated by the results shown in Fig.~\ref{1}.
In the top panel of this figure, we show a contour
plot illustrating the time evolution of the density difference
\begin{equation}
\Delta(x,t) \equiv |u_{\rm num}(x,t)|^2 - |u(x,t)|^2,
\label{D}
\end{equation}
where $u(x,t)$ is the time evolution of the initial condition $u(x,0)=u_b(x)$
(i.e., our approximate analytical solution for the ground state of the system), and $u_{\rm num}(x)$ is the numerically found exact solution of the NLS Eq.~(\ref{PT1}).
Since our approximation in deriving Eqs.~(\ref{TF})-(\ref{tf23})
is valid up to order $\mathcal{O}(\epsilon^2)$,
the deviation from the ``exact'' (numerical) solution should be of order $\mathcal{O}(\varepsilon^2)$. As shown in the 
bottom panel of Fig.~\ref{1}, this is the case indeed:
the maximum of function $\Delta$ stays bounded by a constant prefactor
(of order unity) times $\varepsilon^2$.

%

\subsection{Dark solitons}

\begin{figure}[tbp]
\includegraphics[width=6.5cm]{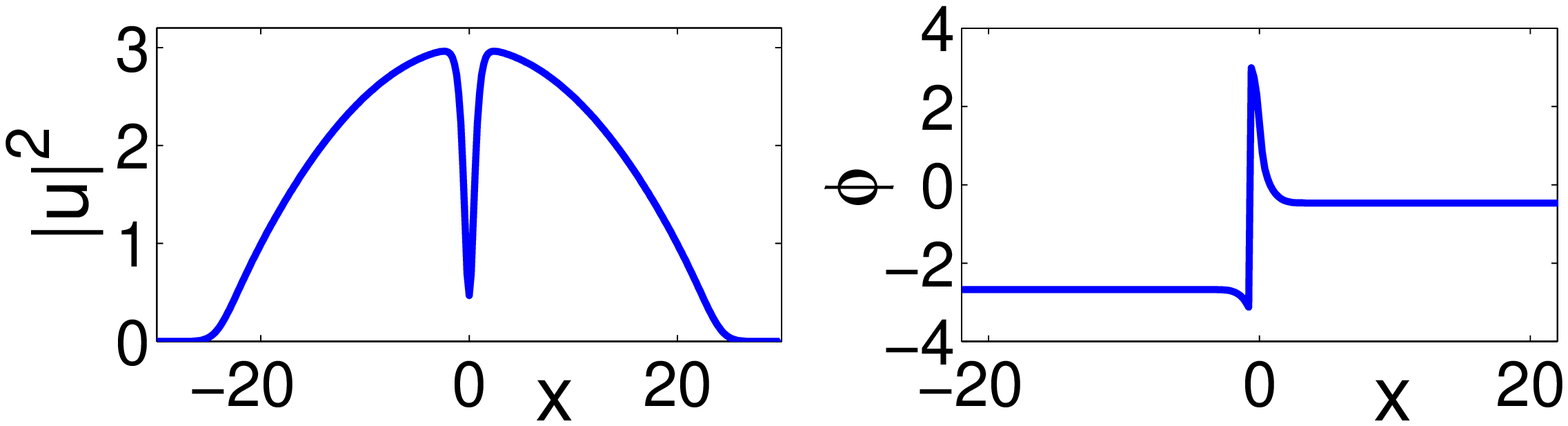}
\includegraphics[width=5.5cm]{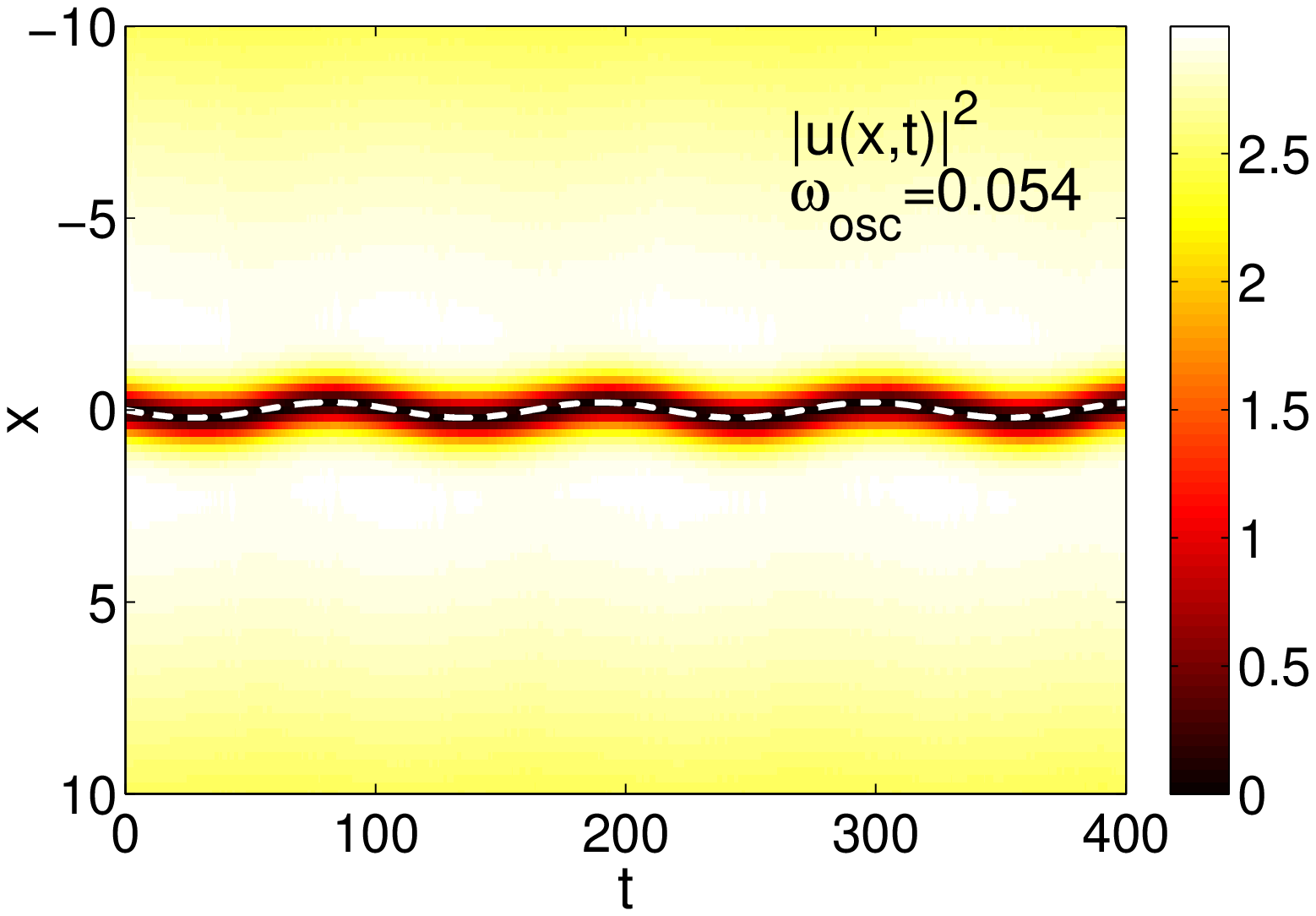}
\caption{(Color online) 
Top: density (left) and phase (right) of a single stationary dark soliton state.
Parameters values are: $\mu=3$, $\Omega=0.1$ and $\varepsilon=0.3$.
Bottom:
contour plot showing the small-amplitude oscillation of a dark soliton for 
$\varepsilon<\varepsilon_{cr}^{(1)}$. The dashed (white) line depicts the 
analytical result of Eq.~(5). 
Parameters are as in the top panel,
but for $\varepsilon=0.04$.}
\label{fig1bb}
\end{figure}

Apart from the ground state, excited states of the system 
---in the form of stationary dark solitons--- can also be found
numerically, by means of the 
Newton's method. In particular, we decompose the field
into the background $u_b$ and the soliton $\upsilon(x,t)$, using the 
product ansatz:
\begin{eqnarray}
\psi=u_b(x)\upsilon(x,t),
\label{dsol}
\end{eqnarray}
%
%
where the function $\upsilon(x,t)$ assumes ---in the absence of the imaginary
potential ($\varepsilon=0$)--- a hyperbolic tangent profile \cite{review,yuri}. 
Then, continuation in $\varepsilon$ results in a dark soliton state, 
such as the one shown in the top panels of Fig.~\ref{fig1bb}.

To describe analytically the dynamics of the dark soliton on top of the TF background, we
substitute Eq.~(\ref{dsol}) into Eq.~(\ref{PT1}) and derive the equation:
\begin{eqnarray}
i\partial_t\upsilon+\frac{1}{2}\partial_{x}^2 \upsilon-|u_b|^2\left(|\upsilon|^2-1\right)\upsilon=-\partial_x      \ln(u_b)\partial_x\upsilon.
\label{pnls}
\end{eqnarray}
Next, we substitute expressions (\ref{TF})-(\ref{tf23}) for $u_b$ and
simplify the resulting equation for $\upsilon(x,t)$
by Taylor expanding the right-hand side term as $\partial_x\ln(u_b)\approx\frac{1}{2}\partial_x f(1+f+f^2+\ldots)$.
Then,
keeping only leading-order terms, up to order $\mathcal{O}(\varepsilon^2)$ [recall that the function $f(x)$ is of order $\mathcal{O}(\varepsilon^2)$],
and using the scale transformations
$t \rightarrow \mu t$ and $x \rightarrow  \sqrt{\mu} x $, we obtain the following
perturbed nonlinear Schr\"{o}dinger (NLS) equation:
\begin{eqnarray}
i\partial_t \upsilon+\frac{1}{2}\partial_{x}^2\upsilon+\upsilon(1-|\upsilon|^2)=\mu^{-2}P(\upsilon),
\label{nlsd}
\end{eqnarray}
where the perturbation $P(\upsilon)$
is given by
%
%
$$
P(\upsilon)\!=\!(1-|\upsilon|^2)\upsilon\left(V+2{\cal W}^2\right)
+ \upsilon_x \left(\frac{1}{2} V_x -2(W-i) {\cal W} \right).
$$
%
%

We now apply the perturbation theory for dark solitons (details can be found in the reviews \cite{review,yuri}).
First we note that in the absence of the perturbation, $P(\upsilon)=0$, Eq.~(\ref{nlsd}) possesses a
dark soliton solution of the form:
\begin{equation}
\upsilon(x,t)=\cos \varphi \tanh \xi +i \sin \varphi,
\label{dark}
\end{equation}
where $\xi \equiv \cos \varphi \left[ x-x_0(t)\right]$, with
$\varphi$ and $x_0=(\sin \varphi)t$ representing the soliton phase angle
and center of the soliton, respectively. Then, in the case $P(\upsilon) \ne 0$,
and in the framework of the adiabatic approximation, the functional form of
the soliton of  Eq.~(\ref{nlsd}) is assumed to be unchanged, but its parameters $\varphi$
and $x_0$ become unknown slowly-varying functions of time. We find that the evolution of these parameters,
which is determined by the perturbation-induced change of the energy of the system \cite{review,yuri},
is described by the following equations:
\begin{eqnarray}
\frac{dx_0}{dt}&=&\sin \varphi(t), \label{dxdt} \\
%
\frac{d\varphi}{dt}&=&-\frac{1}{2}\partial_x V - \int\sech^4(\xi)\left[\tanh(\xi){\cal W}^2
+W {\cal W} \right] dx, \nonumber \\
\label{integ}
\end{eqnarray}
%
%
where we have assumed almost black solitons, with sufficiently small phase angles.
This way, 
we can combine Eqs.~(\ref{dxdt}) and (\ref{integ}) and derive, for a given $W(x)$, 
an equation of motion for the soliton center $x_0$. 
Hereafter, we specify a Gaussian-shaped imaginary potential of the form (\ref{W}).
Note that other choices, e.g., $W(x)=\varepsilon\, {\rm sech}^2(x)
\tanh(x)$, have led to
similar results. To examine the stability of the equilibrium at $x_0=0$,
we Taylor expand Eq.~(\ref{integ}),
obtaining to leading order
%
\begin{equation}
\frac{d^{2}x_{0}}{dt^{2}}= -\omega_{\rm osc}^2\, x_0,
\quad
\omega_{\rm osc}^2 \approx  \left( \frac{\Omega}{\sqrt{2}}\right)^{\!2}-\frac{6}{5} \varepsilon^2.
\label{solev}
\end{equation}
Equation~(\ref{solev}) implies that if the amplitude $\varepsilon$ of $W(x)$ is less than a critical value,
$\varepsilon_{cr}^{(1)}=\sqrt{5/12}\,\Omega$, then the soliton performs oscillations in the
complex potential with frequency $\omega_{\rm osc}$.
Such a case is demonstrated in the bottom panel of Fig.~\ref{fig1bb}, where we show
a dark soliton oscillating around the trap center for $\varepsilon=0.04<\varepsilon_{cr}^{(1)}$.
The numerically found trajectory, obtained by direct numerical integration of Eq.~(\ref{PT1}), is compared with the analytical result of Eq.~(\ref{solev}) [dashed (white) line]; as is observed, the agreement between the two is excellent.

\begin{figure}[tbp]
\includegraphics[width=7cm]{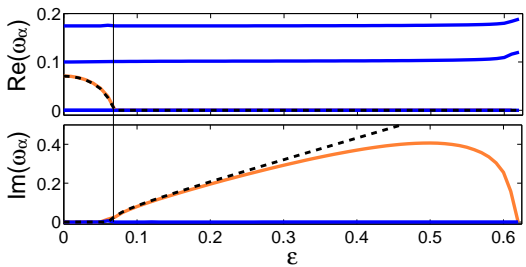}
\caption{(Color online)
The linear spectrum, as obtained numerically by the BdG analysis,
of the single dark soliton
branch for $\Omega=0.1$ and $\mu=3$:
top (bottom) panel shows the real (imaginary) part of the lowest eigenfrequencies $\omega$
as a function of the amplitude $\varepsilon$.
The lowest solid (orange) line in the top panel and the upper solid (orange) curve in the bottom panel depict, respectively, the real and imaginary part of the anomalous mode eigenfrequency $\omega_\alpha$. The dashed (black) lines in both panels indicate the analytical result of Eq.~(\ref{solev}). 
The thin vertical line shows the point $\varepsilon_{cr}^{(1)}$, where the 
anomalous mode eigenfrequency becomes imaginary.}
%
\label{fig2cc}
\end{figure}

On the other hand, Eq.~(\ref{solev}) dictates that if $\varepsilon>\varepsilon_{cr}^{(1)}$ then the soliton will become unstable. The above prediction has been confirmed numerically, both by means of direct simulations and by employing a BdG analysis. In the framework of the latter, the stability of the dark soliton is studied by considering the
anomalous mode eigenfrequency $\omega_\alpha$ (which 
is associated with  the dark soliton motion~\cite{theocharis,review}). 
If the imaginary part of this eigenfrequency is zero (nonzero) then the soliton is stable (unstable).
We have found that this eigenfrequency is real for $\varepsilon<\varepsilon_{cr}^{(1)}$ and, in this case, $\omega_\alpha$ coincides with the analytically found oscillation frequency $\omega_{\rm osc}$. On the other hand, 
for values $\varepsilon>\varepsilon_{cr}^{(1)}$, the anomalous mode eigenfrequency $\omega_\alpha$
becomes imaginary,
thus signaling the onset of the spontaneous symmetry-breaking (SSB) 
instability of the dark soliton, which displaces the
dark soliton from the trap center.
The detailed dependence of $\omega_\alpha$ on the amplitude $\varepsilon$ of
the imaginary potential $W$, as found by the BdG analysis,  is
illustrated in
Fig.~\ref{fig2cc}. It is observed that
the anomalous mode $\omega_\alpha$ initially moves towards the
origin, and past the critical point $\varepsilon_{cr}^{(1)}$ (cf.~thin vertical line),
becomes imaginary, 
manifesting the soliton's exponential instability.
Importantly, as shown in
Fig.~\ref{fig2cc},
for small $\varepsilon$ the agreement between the analytical
prediction of Eq.~(\ref{solev}) [dashed (black) line] and the BdG
numerical result [lowest solid (orange) line] is excellent.
\subsection{Effect of noise-induced perturbations}

Let us now consider an experimentally relevant situation, where the (localized) gain 
which acts on the system is associated with the presence of noise. 
This is important in order to ensure the robustness of our results
presented above in realistic cases where  $\mathcal{PT}$-symmetry
is not strictly enforced.
In such a case, an important 
question is if and how the noise-induced perturbation affects the stability and dynamics 
of the ground state and the dark soliton. To address this question, we assume that 
---in the simplest approximation--- the gain side of the potential $W(x)$ (for $x>0$) 
now becomes $W(x)[1+\delta\, n(x)]$; here, $n(x)$ is a uniformly distributed noise of amplitude 
$\delta$. 
%
%
Thus, generally, one expects from Eq.~(\ref{dNdt})
that since the noise $n(x)$ is not parity symmetric, 
$dN/dt \ne 0$, i.e., for any time instant $t$, the system will either 
grow ($dN/dt>0$) or decay ($dN/dt<0$). As a result, stationary states (such as 
the TF background ground state or a single- or multiple-dark-soliton state) cannot generically exist, at least in the case of relatively large noise amplitude.

\begin{figure}[tbp]
 \includegraphics[width=4.2cm]{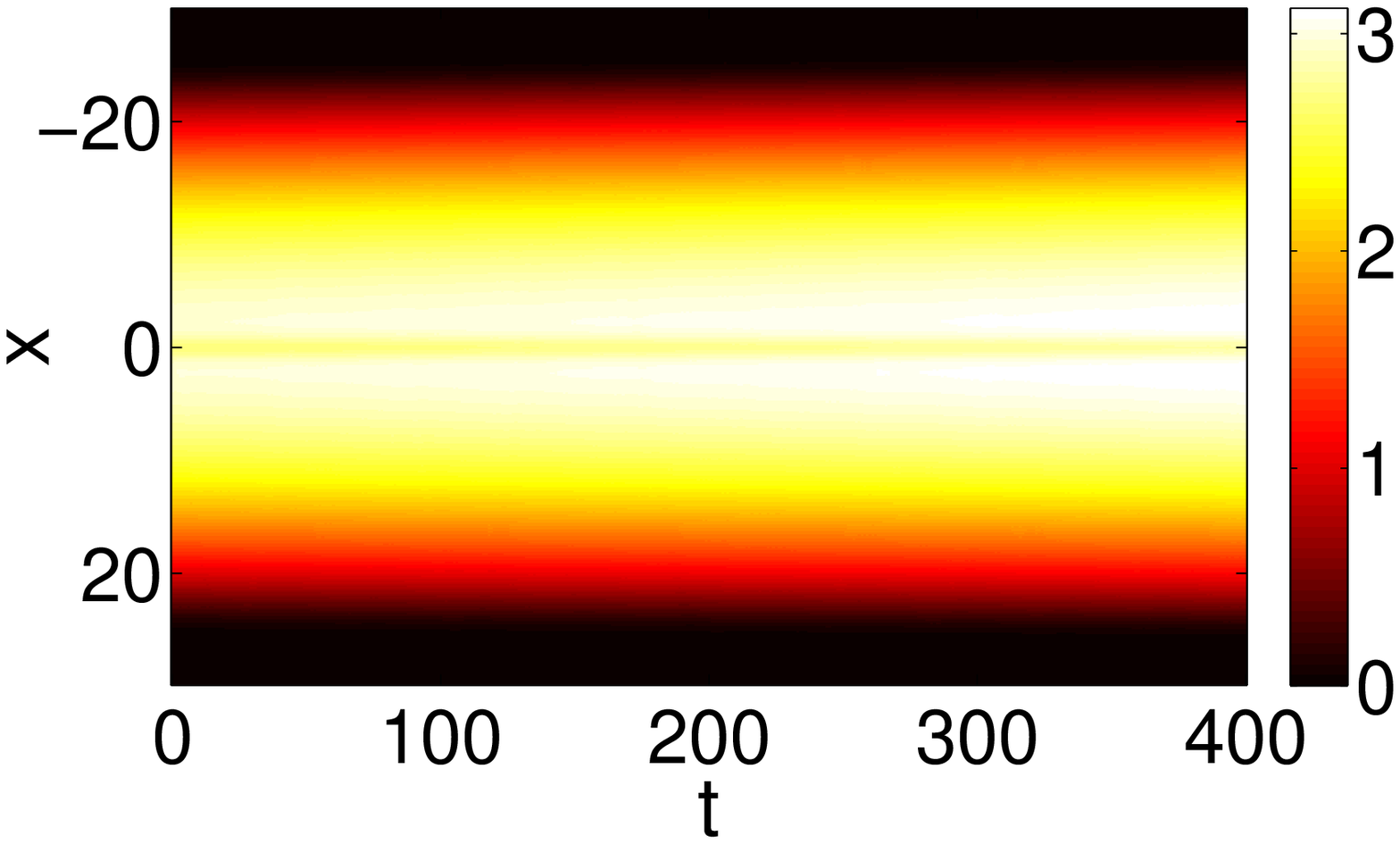}
~\includegraphics[width=4.2cm]{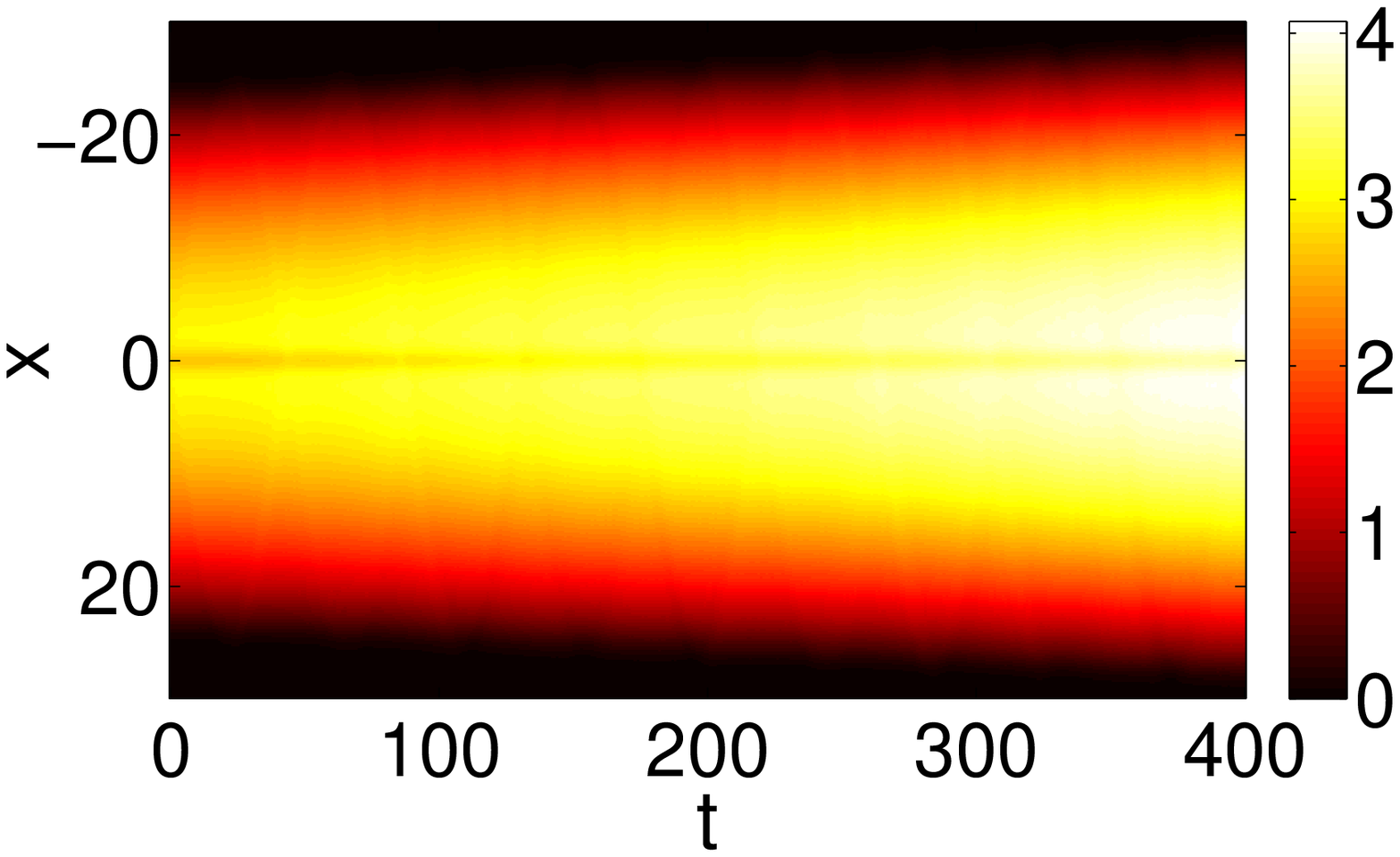}
\\[1.0ex]
 \includegraphics[width=4.2cm]{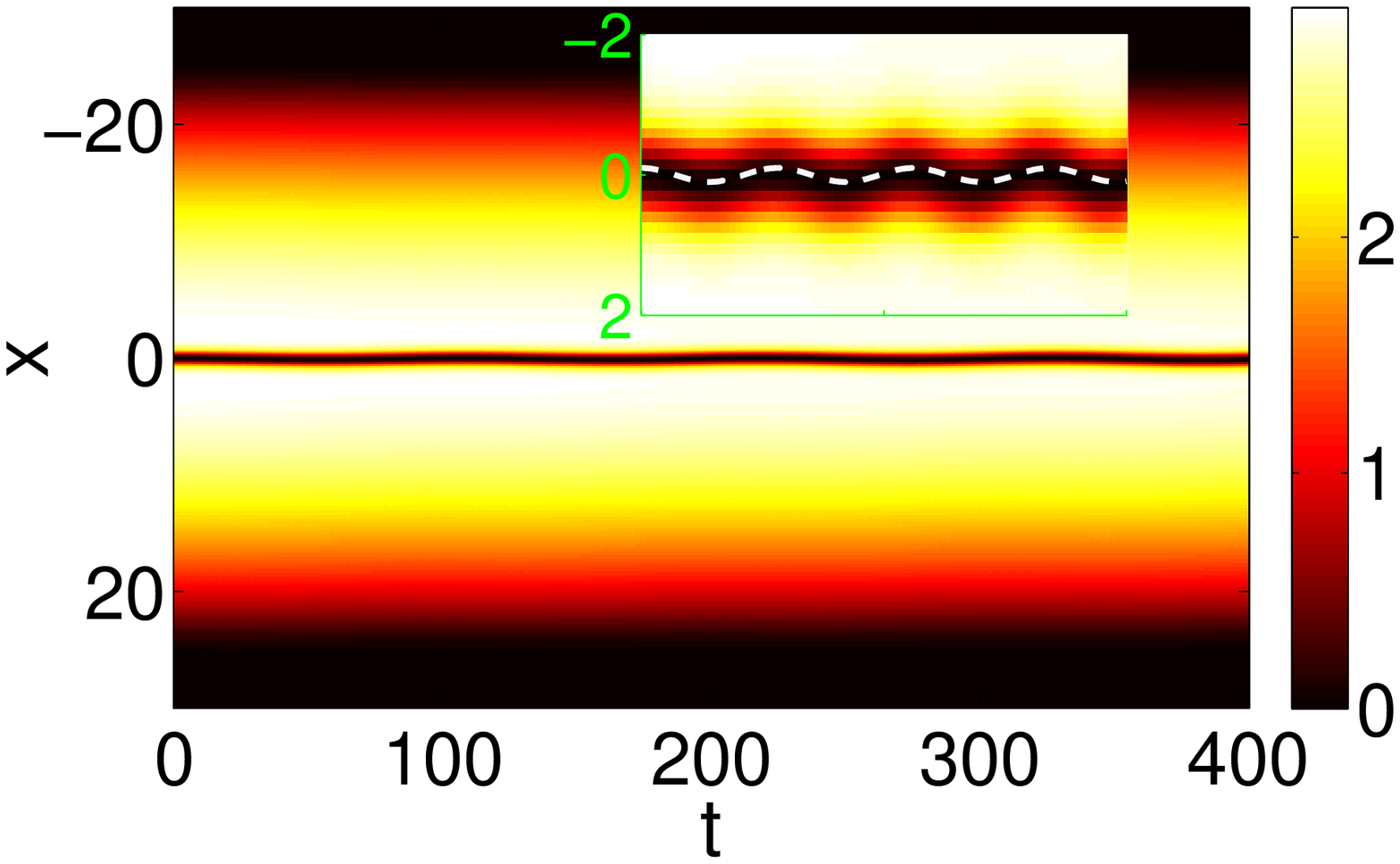}
~\includegraphics[width=4.2cm]{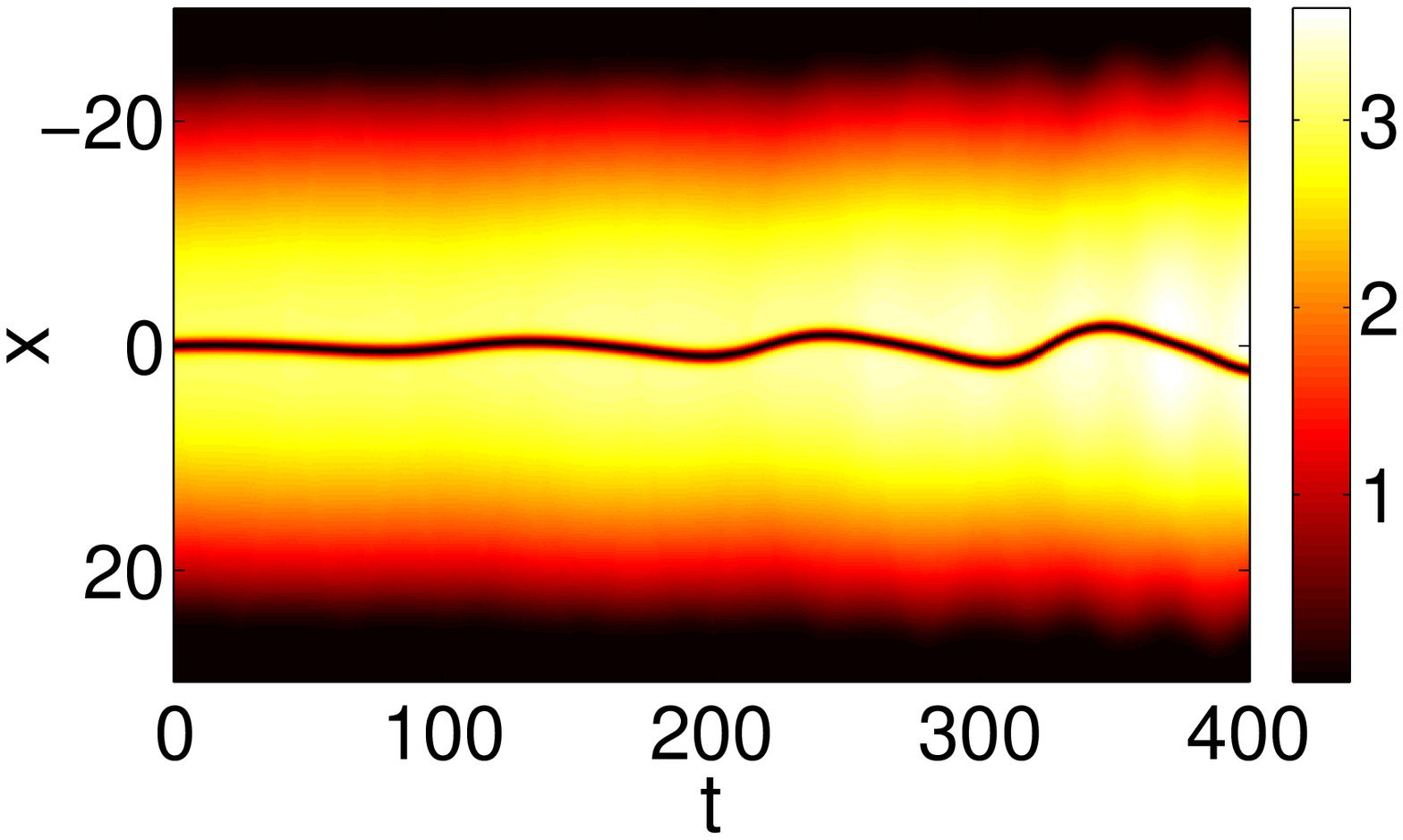}
\caption{(Color online) Contour plots, showing the evolution of the density 
when noise is added to Eq.~(\ref{PT1}), on the gain side of the imaginary potential ($x>0$).
Top: for a relatively small amplitude noise, $\delta=0.1$,  the TF background evolves practically unaffected [left], while for a larger noise, $\delta=1$ the background grows [right]. In both top panels $\varepsilon=0.4$. 
Bottom: a dark soliton, performs small amplitude oscillations (see inset) after 
perturbed by a small amplitude noise of $\delta=0.1$ [left], while a dark soliton 
is oscillating with a growing amplitude for a larger amplitude noise, 
$\delta=1$ [right]. The amplitude of the imaginary potential is $\varepsilon=0.04$, 
for both panels, where in the absence of noise the soliton is stable 
[cf. Fig.~(\ref{fig2cc})]. Other parameter values are: $\mu=3$, $\Omega=0.1$.}
\label{fig2dd}
\end{figure}

To investigate the significance of this effect of noise, we have 
numerically integrated Eq.~(\ref{PT1}) with prototypical
initial conditions of relevance to our study including 
the TF background and the dark soliton,
and have let the system to evolve in the presence of noise perturbations. 
The results of our simulations can be summarized in the examples shown in 
Fig.~\ref{fig2dd}, where contour plots showing the evolution of the density of the TF background (top panels) 
and a single dark soliton (bottom panels) are given. It is observed that if the noise amplitude $\delta$ is sufficiently 
small (the value $\delta=0.1$ was used in the left panels of the figure) then the dynamics is practically unaffected 
by the effect of noise (in fact, the effect of noise can be observed for larger values of $\delta$, as is explained below). Note that in the case of the dark soliton, the noise induces the soliton to perform
small-amplitude oscillations (cf.~inset in the bottom left panel of Fig.~\ref{fig2dd}) which can be very well 
described by Eq.~(\ref{solev}). On the other hand, if the noise is strong enough (as, e.g., with $\delta=1$ used 
in the right panels) then the ground state either grows or decays, 
depending on the initial sign of $dN/dt$ 
(which depends, in turn, on the particular noise realization), with an average growth rate determined by the parameter $\delta$.
Notice that, in this case of large-amplitude noise, the dark soliton 
becomes thermodynamically unstable and is 
displaced from the center, performing oscillations of growing amplitude.  The 
approximate evolution equation for the dark soliton, Eq.~(\ref{solev}), is 
not valid here, since the noise term becomes of the same order 
$\varepsilon$ as the rest of the perturbation and needs to be explicitly
accounted for in the anti-damped dynamics of the solitary wave.

Overall, these results support that under weak noise perturbations,
the phenomenology presented above (and below) will persist. Yet, for
strong random perturbations, the phenomenology changes considerably
and should be considered separately in further detail.

\section{
Nonlinear $\mathcal{PT}$ phase transitions}
\begin{figure}[tbp]
\includegraphics[width=7.0cm]{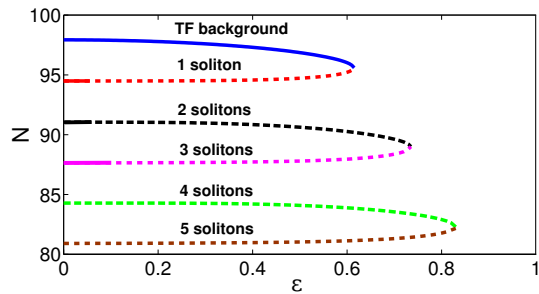}
\caption{(Color online)
The power $N$ as function of the strength $\varepsilon$ of the imaginary potential,
depicting the full bifurcation diagram for the ground state and excited states in 1D.
The diagram encompasses the pairwise blue-sky
bifurcations/disappearances of the nonlinear states, namely the
ground state with the 1st excited state (single dark soliton), the
2nd with the 3rd excited state (two- and three-dark-solitons), the
4th with the 5th, and so on.
Solid (dashed) lines indicate dynamically stable (unstable) branches. Here,
$\mu=3$ and $\Omega=0.1$.}
\label{fig3}
\end{figure}

Let us now return to the results of the BdG analysis presented in the previous section,
and discuss in more detail what happens beyond the SSB instability of the single dark soliton state.
As shown in Fig.~\ref{fig2cc}, for
values of $\varepsilon>\varepsilon_{cr}^{(1)}$
the unstable imaginary eigenvalue
makes a maximal excursion along the imaginary line and returns to
the origin at a second critical point, $\varepsilon_{cr}^{(2)}=0.62$,
finally colliding with it. The branch of single soliton solutions
disappears past this critical point.
To better understand how the branch ceases to exist, we first
observe 
(see top panel of Fig.~\ref{fig1bb}) that
the density profile of the soliton becomes increasingly shallower
(i.e., more ``grey'')
as $\varepsilon$ grows and the second critical point is approached.
This is due to the development of an increasingly strong {\it even}
imaginary part of the solution.
Furthermore, the
stable background (ground state) solution $u_b(x)$
[cf. Eqs.~(\ref{TF})-(\ref{tf23}) and top panel of Fig.~\ref{fig0}]
develops an {\it odd} imaginary part resembling a (progressively darker) grey soliton.
Finally, at $\varepsilon=\varepsilon_{cr}^{(2)}$, the profiles of these modes become
identical and disappear in a blue-sky bifurcation through their collision.
This is shown in
Fig.~\ref{fig3}, where the power $N$
is shown as a function of $\varepsilon$.
The top solid (blue) branch
shows the stable ground state, $u_b$, which ultimately collides with the one soliton
(first excited) state at $\varepsilon\approx 0.62$ (for $\mu=3$ and $\Omega=0.1$).

Importantly, we have confirmed that the above description holds
also for higher excited states (multiple dark soliton solutions), as shown in
Fig.~\ref{fig3}: each pair of the higher excited states (2nd with 3rd,
4th with 5th, etc.) also disappears in a blue-sky bifurcation.
A general remark is that higher excited states bifurcate for larger values of $\varepsilon$.
Remarkably, this can be thought of as a {\it nonlinear analogue} of the $\mathcal{PT}$ transition,
in analogy with the
pairwise collisions in Ref.~\cite{bend1} (see, e.g., Fig.~1 of that
reference) for the linear setting \cite{note2}.

\begin{figure}[t]
 \includegraphics[width=4.2cm]{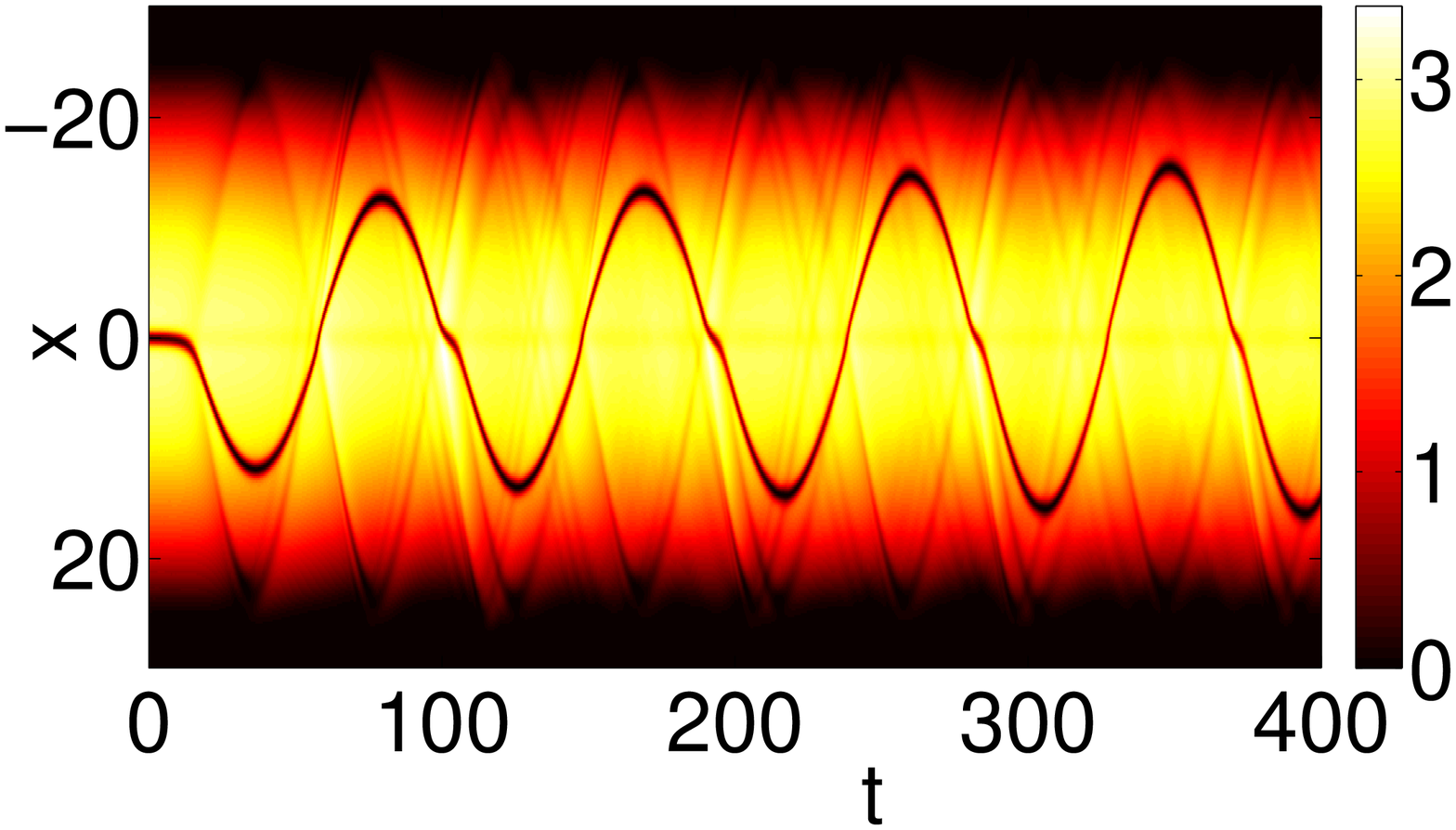}
~\includegraphics[width=4.2cm]{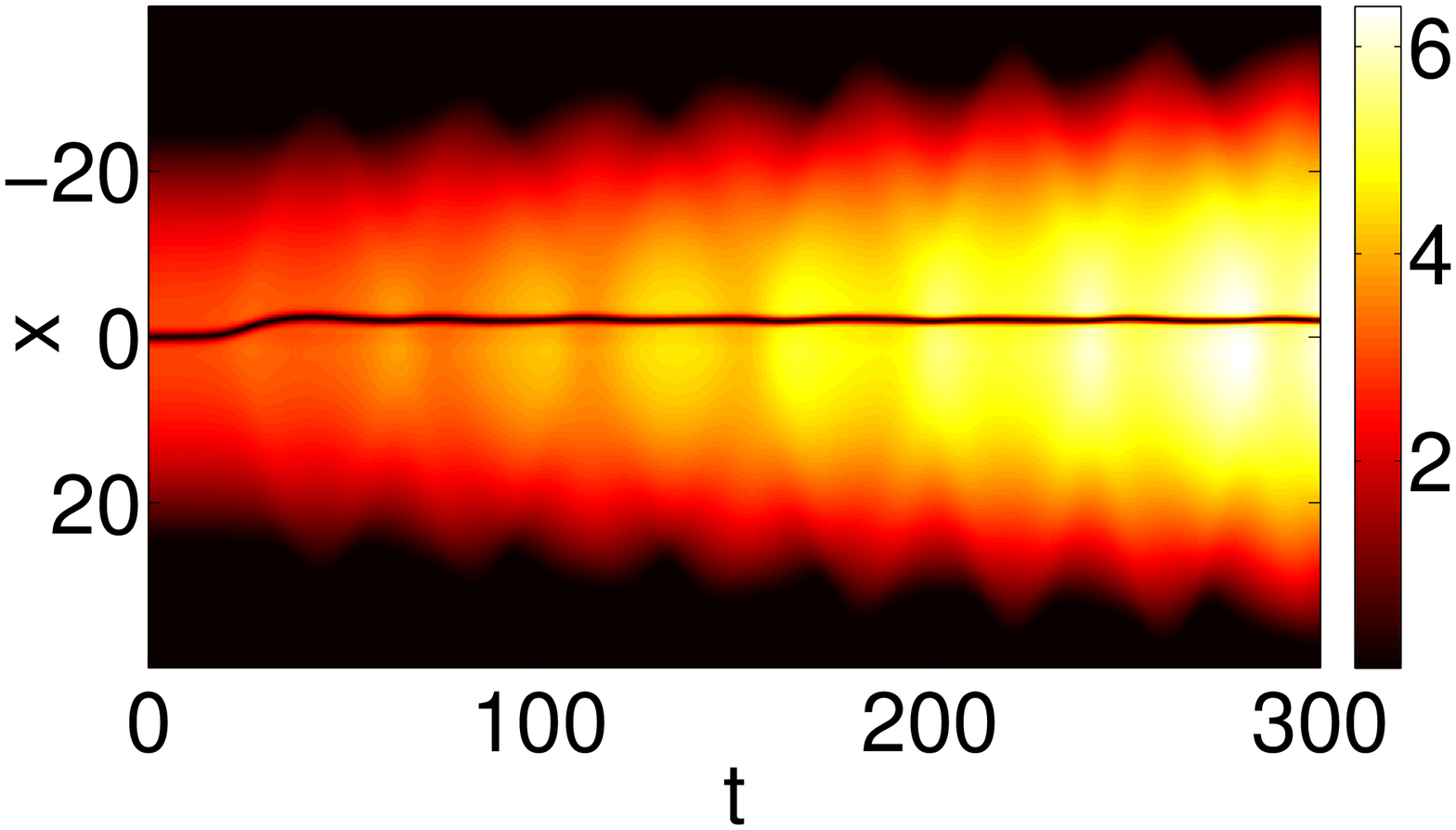}
\\[1.0ex]
 \includegraphics[width=4.2cm]{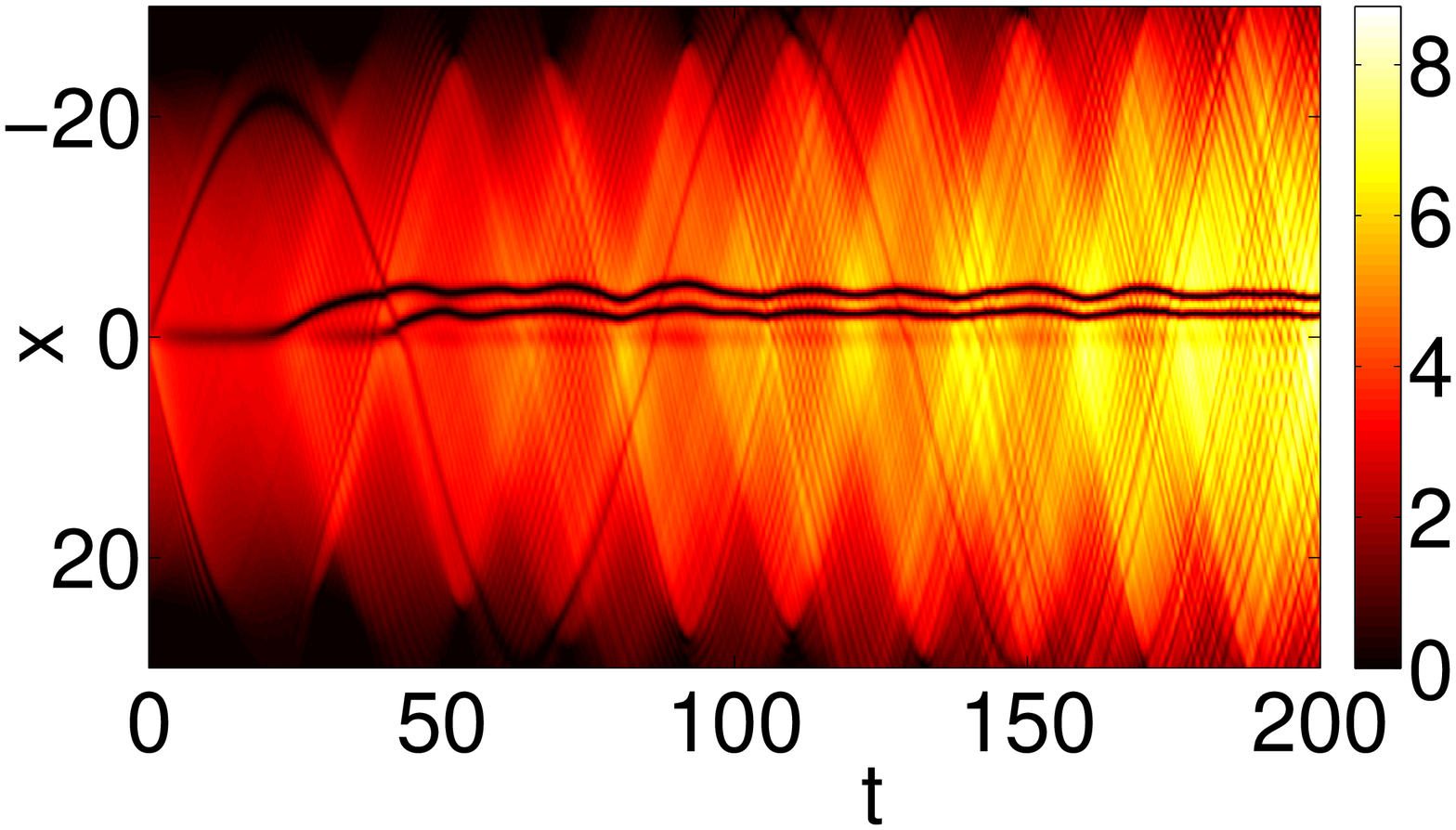}
~\includegraphics[width=4.2cm]{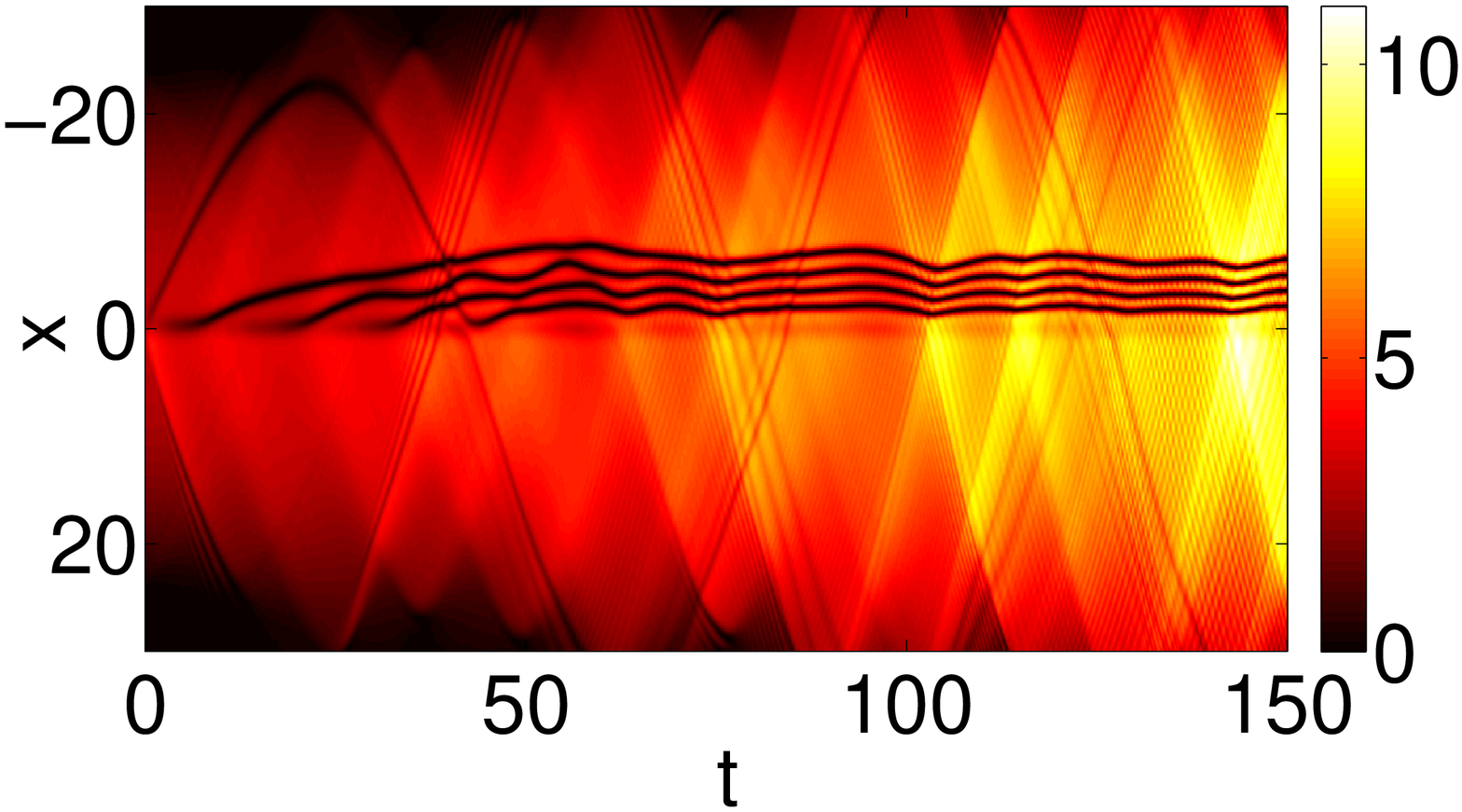}
\caption{(Color online) Bifurcation-induced dynamics. Top panels:
  manifestations of the SSB destabilization scenarios for an unstable dark
  soliton past  $\varepsilon=\varepsilon_{cr}^{(1)}$. Bottom panels:
soliton
generator, spontaneously leading to two or four solitons from
the ground state used for $\varepsilon>\varepsilon_{cr}^{(2)}$.
%
The parameters are $\mu=3$ and $\Omega=0.1$ and
$\varepsilon=0.3$ (top row),
$\varepsilon=0.64$ (bottom left), and
$\varepsilon=0.7$ (bottom right).}
%
\label{fig4}
\end{figure}

A relevant and interesting question concerns the dynamics of the nonlinear 
states when subject to these (SSB and blue-sky) bifurcations.
To answer this, we numerically integrated
Eq.~(\ref{PT1}) and the relevant results are shown in Fig.~\ref{fig4}. In the top panels, we
illustrate the dynamics of the dark soliton upon its destabilization at $\varepsilon=\varepsilon_{cr}^{(1)}$.
When the SSB is manifested, the soliton is either
spontaneously ejected towards the lossy side (and typically found to localize
therein, while the background grows in amplitude and widens) or moves
to the gain side, executing oscillations thereafter. 

On the other hand,
past $\varepsilon=\varepsilon_{cr}^{(2)}$,
using, as an initial condition the form of the TF background (bottom
panels of Fig.~\ref{fig4}),
we have found that a dark soliton train is
spontaneously formed, with an increasingly larger number of solitons as larger values
of $\varepsilon$ are used.
This can be intuitively connected to the observation of Fig.~\ref{fig4}
that higher excited multi-soliton states persist for larger 
$\varepsilon$ than lower ones.
Again, it is typically observed that the solitons are nucleated and stay in the vicinity of the
global minimum of $W(x)$, which corresponds to the ``lossy'' side of the imaginary potential.

\section{Two-dimensional Generalizations}\label{sec:2D}

\begin{figure}[tbp]
\includegraphics[width=3.80cm]{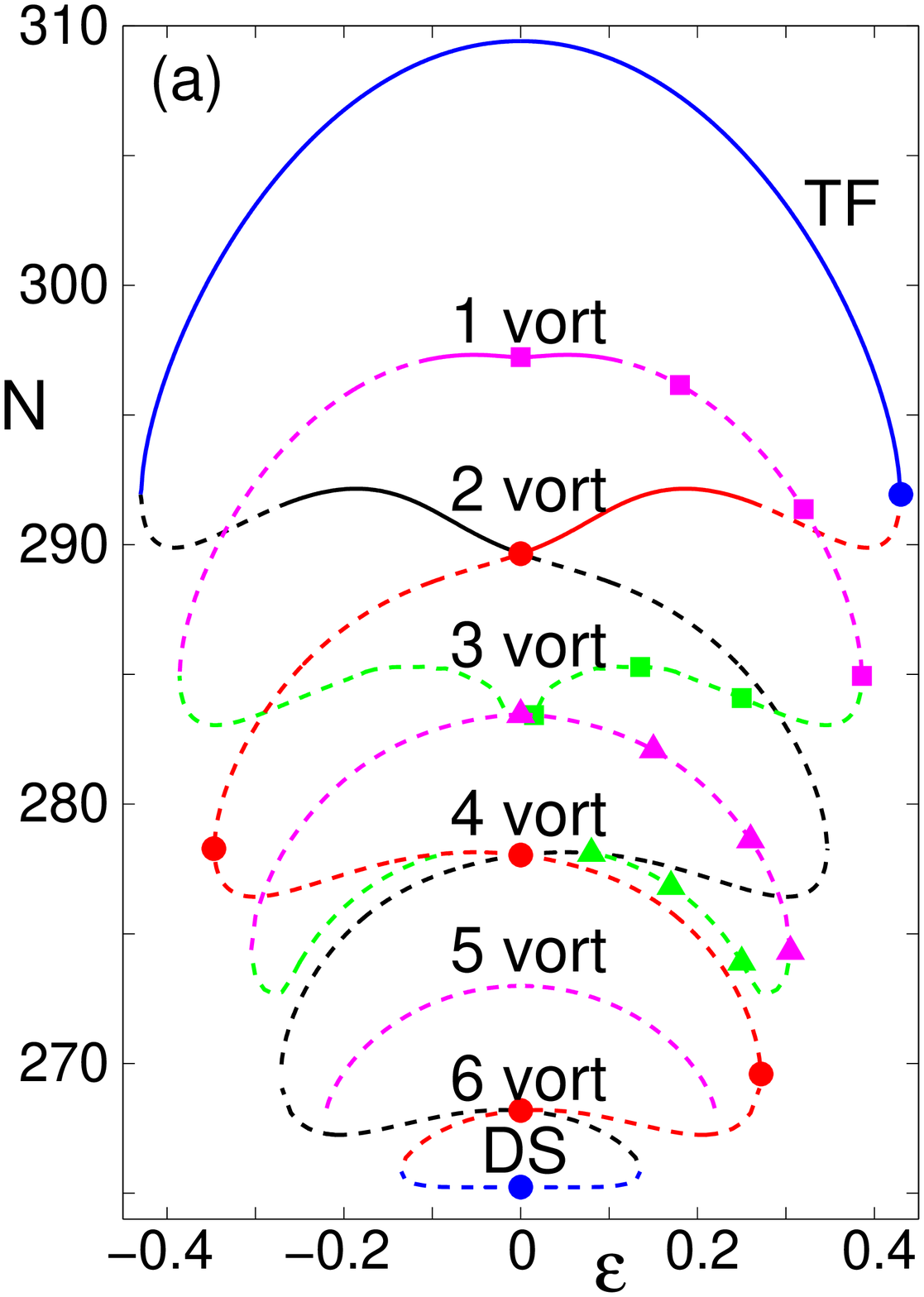}
\includegraphics[width=1.51cm]{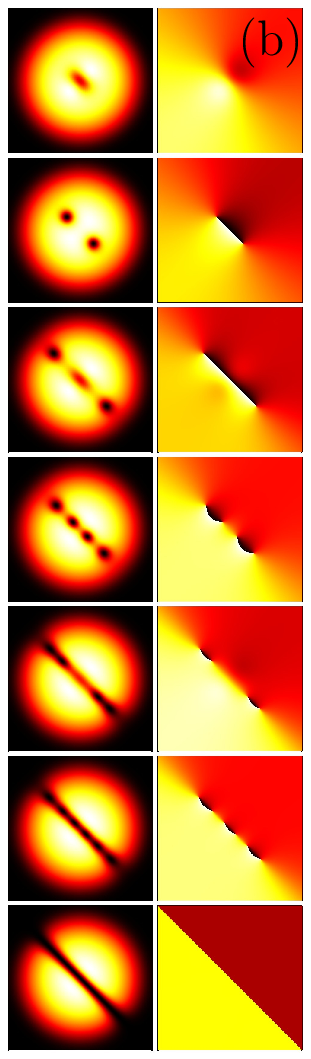}
\includegraphics[width=1.51cm]{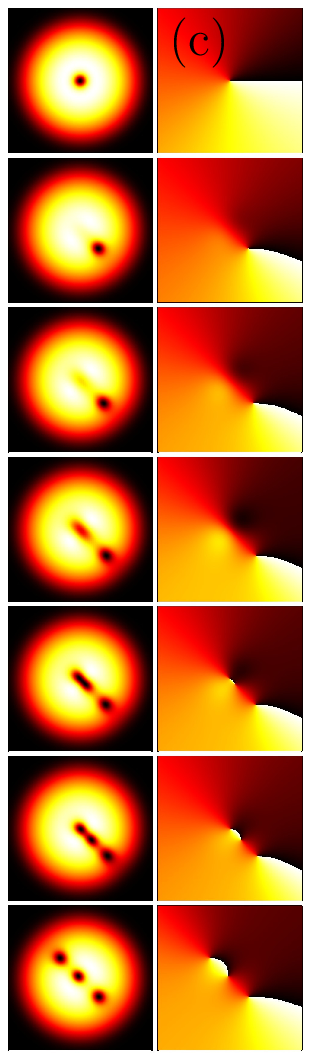}
\includegraphics[width=1.51cm]{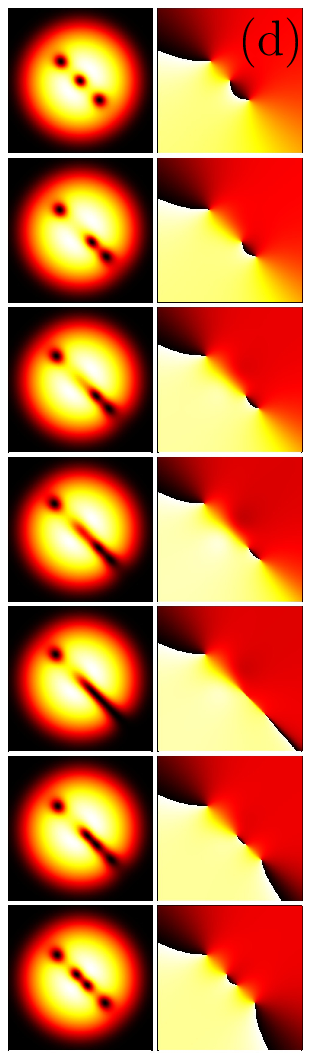}
\caption{(Color online)  The 2D generalization.
(a) Bifurcation diagram for the 2D stationary nonlinear (vortex and
DS stripe) states. Stable (unstable) branches,
as per the corresponding BdG analysis,
are depicted with solid (dashed) lines.
(b) Series of density (left) and phase (right) configurations
along the branch with even number of vortices corresponding to
the circles in panel (a) [from top to bottom].
(c) Same as (b) for the branch starting with one vortex
and connecting with three vortices corresponding to the squares
in panel (a) [from top to bottom].
(d) Same as (b) for the branch starting with three symmetric
vortices and ending with four vortices corresponding to the triangles
in panel (a) [from top to bottom].
Parameter values are: $\mu=2$ and $\Omega=0.2$.
The field of view for the configurations is $[-10.5,10.5]\times[-10.5,10.5]$.
}
\label{fig5}
\end{figure}

Finally, we consider the case of a 2D $\mathcal{PT}$-symmetric potential
with a real parabolic part
$$
V(x,y)=\frac{1}{2}\Omega^2 (x^2+y^2),
$$
and an odd [$W(-x,-y)=-W(x,y)$] imaginary part 
$$
W(x,y)=\varepsilon (x+y) e^{-(x^2+y^2)/4}.
$$
The bifurcation of the nonlinear structures emerging in 2D follows a
similar, but also more complex, pattern than in the corresponding 1D setting.
Figure \ref{fig5} depicts the full bifurcation scenario for solutions
bearing no vortices (the TF background cloud), one to six vortices,
and the dark soliton stripe.
As in 1D, the TF background is stable in all its domain of existence and
collides, in a blue-sky bifurcation, for a large enough value of $\varepsilon$,
with an excited state. However, in contrast to the 1D case where this
collision happens with the first excited state, in 2D the
collision occurs with the {\em second} excited state, due to the
absence
of net topological charge in such a vortex-dipole (see top-right red
curve) bearing two opposite charge vortices emerging from
the central dip of the TF background.
At this critical point $\varepsilon=\varepsilon_{cr}^{(2)}$
the dipole branch is unstable,
having been destabilized
through an SSB bifurcation at
an $\varepsilon=\varepsilon_{cr}^{(1)}>0$ value
(below which for  $\varepsilon>0$ the dipole
is stable ---see portion of red solid line in the figure).
As this branch is followed (from top to bottom in the figure),
a series of bifurcations occur where the existing vortices
are drawn to the periphery of the cloud, a dip in the center
deepens leading eventually to a new vortex pair emerging (i.e., a
higher excited state).
In this manner the branches with {\em even} number of vortices
are all connected. As more and more vortex pairs emerge, the
cloud ``saturates'' and can no longer fit in new vortex pairs
finally colliding with a dark soliton stripe (see lower
blue branch in the figure).
This overall bifurcating structure of even vortex numbers
---with a $\varepsilon \rightarrow -\varepsilon$
symmetry where the solutions are just flipped by $(x,y) \rightarrow (-x,-y)$---
is depicted, with density and phase profiles, in the
series of panels of Fig.~\ref{fig5}(b).

As for the bifurcation scenario of {\em odd} number of vortices,
the first excited state bearing a single vortex
at the origin (for $\varepsilon=0$) is stable for small values of
$\varepsilon$, while it again sustains an SSB bifurcation for larger
$\varepsilon$.
As $\varepsilon$ increases the vortex moves towards
the periphery of the cloud and a dip at the center of the cloud
deepens until a vortex pair emerges from it. This scenario
connects the one-vortex branch with the {\em asymmetric} three-vortex
($+ - +$ vortex tripole) branch,
as it is depicted with the top (magenta and green) lines in panel \ref{fig5}(a)
and the series of snapshots in panels \ref{fig5}(c).
As it is evident from the figure,
the asymmetric three-vortex branch eventually connects with the
symmetric one for values of $\varepsilon \rightarrow 0$.
A similar bifurcation occurs with the symmetric three-vortex
branch, which becomes asymmetric with a deepening dip at the
center where a vortex pair emerges (at the same time that a
vortex is lost at the periphery),
connecting in this way with the
four-vortex branch [see series of snapshots in panels \ref{fig5}(d)].

As for the dynamics of unstable steady states, we have observed
---in analogy with the 1D case--- that (a) a single vortex tends to migrate
towards the minimum of the lossy side of the potential, while the
remaining vortices (if present) perform almost circular orbits at the
periphery of the cloud where they are eventually absorbed;
%
and (b) past $\varepsilon=\varepsilon_{cr}^{(2)}$,
using as an initial condition the form of the TF background,
also produces the spontaneous formation of an increasing
number of vortices for larger values of $\varepsilon$
(namely, a ``vortex
generator'').
%
%
It is worth mentioning that the precise structure of the bifurcation
diagram depends of the values of the propagation constant $\mu$ and
the trap strength $\Omega$. For weaker $\Omega$ and/or larger $\mu$
the extent of the TF background will be larger allowing for a
longer bifurcating chain of higher-order vortex states.
%
Nonetheless, the displayed SSB instabilities and phenomenology and  the
nonlinear $\mathcal{PT}$ transition involving the cascade of blue-sky
bifurcations
(notice that in the 2D case the order is reversed and the largest
$\varepsilon$
bifurcation is that involving the TF and the dipole states)
appear to be universal in confining $\mathcal{PT}$-symmetric
potentials.



\section{Conclusions}

In the present work, we have developed some fundamental
insights stemming from the interplay of defocusing nonlinearity
and $\mathcal{PT}$-symmetric confining potentials.
We identified both a symmetry-breaking
bifurcation destabilizing the dark solitons that leads to non-stationary
dynamics, as well as a nonlinear analogue of the $\mathcal{PT}$ transition that
eventually terminates both the ground state and the dark soliton
branch, yielding purely gain-loss dynamics within the system.
Similar bifurcation
phenomena and dynamics of mobility or of spontaneous emergence of
dynamical patterns forming out of the destabilization of the nonlinear states
were identified in two-dimensional settings, for
vortices. 

These investigations, we believe,
pave the way for studying $\mathcal{PT}$-symmetric systems in the context of
defocusing  nonlinearities  and of higher dimensional systems,
which are some of the natural extensions of the $\mathcal{PT}$-symmetric literature.
A canonical set of investigations which is still missing concerns the effects
of such potentials in three-dimensional continuum or higher
dimensional lattice contexts, as well as the manipulation of nonlinear states
emerging in these systems. 
Another relevant possibility arising from our considerations here in
is the study of inexact $\mathcal{PT}$-symmetric 
nonlinear systems and their comparison to exact ones, as well
as the consideration of the interplay of the nonlinearity with
merely loss (rather than gain-loss) in the passive-$\mathcal{PT}$ nonlinear
settings.
These themes will be pursued in future works.

\section*{Acknowledgments} 
The work of D.J.F. was partially supported by the Special 
Account for Research Grants of the University of Athens.
P.G.K. gratefully acknowledges support from the National Science
Foundation under grants DMS-0806762 and CMMI-1000337, as well as
from the Alexander von Humboldt Foundation, the Alexander S. Onassis
Public Benefit Foundation and the Binational Science Foundation.
R.C.G.~gratefully acknowledges support from the National Science Foundation
under grant DMS-0806762.



\begin{thebibliography}{99}
\bibitem{note1} Note that $\varepsilon$ is measured in units of the free space wavelength
(in optics) or in units of energy (in BECs).
\bibitem{note2} Despite the similarity of
the two figures, it does not escape us that an interesting 
difference exists: in the linear picture the lowest eigenvalues
collide the last, while in the 1D nonlinear case the lowest eigenstates disappear the 
first, a feature that important consequences for the supercritical dynamics of Fig.~\ref{fig3}.
\end{thebibliography}
\end{document}